\documentclass[traditabstract]{aa} % for the abstract without structuration 
\usepackage{graphicx}
\usepackage{natbib}
\bibpunct{(}{)}{;}{a}{}{,}
\usepackage{txfonts}
\usepackage[scientific-notation=true]{siunitx}
\usepackage{color}

\begin{document}

   \title{Cosmic variance in [O/Fe] in the Galactic disk}
%   \subtitle{}
   \author{S. Bertran de Lis
                \inst{1,} \inst{2}
   \and
                C. Allende Prieto
            \inst{1,} \inst{2}
   \and 
       S. R. Majewski
       \inst{3}     
   \and
                R. P. Schiavon
                \inst{4}            
        \and
                J. A. Holtzman
                \inst{5}    
   \and
                M. Shetrone
                \inst{6}            
   \and
        R. Carrera
            \inst{1,} \inst{2}
   \and
                A. E. Garc\'{\i}a P\'erez       
                \inst{1,} \inst{2}
   \and    
       Sz.~M{\'e}sz{\'a}ros
       \inst{7}  
   \and
                P. M. Frinchaboy
                \inst{8}
   \and
                F. R. Hearty
                \inst{9}
   \and
                D. L. Nidever
                \inst{10,} \inst{11}
   \and
                G. Zasowski
                \inst{12}
   \and
       J. Ge
       \inst{13}       
          }

   \institute{
                Instituto de Astrof\'{\i}sica de Canarias, V\'{\i}a L\'actea, 38205 La Laguna, Tenerife, Spain        
   \and
                Universidad de La Laguna, Departamento de Astrof\'{\i}sica, 38206 La Laguna, Tenerife, Spain 
   \and     
          Department of Astronomy, University of Virginia, Charlottesville, VA 22904-4325, USA      % Majewski
   \and
                  Astrophysics Research Institute, Liverpool John Moores University, Liverpool, L3 5RF, UK % Schiavon
   \and      
                New Mexico State University, Las Cruces, NM 88003, USA          % Holtzman
   \and
         University of Texas at Austin, McDonald Observatory, Fort Davis, TX 79734, USA % Shetrone
   \and
         ELTE Gothard Astrophysical Observatory, H-9704 Szombathely, Szent Imre herceg st. 112, Hungary%Meszaros
   \and
                Texas Christian University, Fort Worth, TX 76129, USA   % Frinchaboy
   \and
        405 Davey Laboratory, Pennsylvania State University, University Park PA % Hearty
   \and
                Large Synoptic Survey Telescope, 950 North Cherry Ave, Tucson, AZ 85719  % Nidever
   \and
                Steward Observatory, 933 North Cherry Ave, Tucson, AZ 85719                    % Nidever
   \and
                  Johns Hopkins University, Baltimore, MD, 21218, USA  % Zasowski
   \and
          Department of Astronomy, University of Florida, Bryant Space Science Center, Gainesville, FL, 32611-2055, USA
       %      \thanks{}
             }

   \date{Received 24 November 2015 / Accepted 8 March 2016}

\abstract{
We examine the distribution of the [O/Fe] abundance ratio in stars across the Galactic disk
using H-band spectra from the Apache Point Galactic Evolution Experiment (APOGEE). We minimize
systematic errors by considering groups of stars with similar atmospheric parameters. 
The APOGEE measurements in the Sloan Digital Sky Survey Data Release 12 reveal 
that the square root of the star-to-star cosmic variance in the oxygen-to-iron ratio at a given metallicity 
is about  0.03--0.04 dex in both the 
thin and thick disk. This is about twice as high as the spread found for solar
twins in the immediate solar neighborhood and the difference is probably associated to the wider range of galactocentric distances spanned by APOGEE stars. We quantify the uncertainties by examining the spread among 
stars with the same parameters in clusters; these errors are a function of effective temperature 
and metallicity, ranging between 0.005 dex at 4000 K and solar metallicity, 
to about 0.03 dex at 4500 K and [Fe/H]$\simeq -0.6$.
We argue that measuring the spread in [O/Fe] and other abundance ratios provides strong
constraints for models of Galactic chemical evolution.
}


   \maketitle
%
%________________________________________________________________

\section{Introduction}

Cosmic variance in the chemical composition of stars in a galaxy is a natural 
outcome of the discrete character of the production and return of nucleosynthetic 
yields from preceding generations of stars to the interstellar medium (ISM). 
At a given location within a galaxy, slower supernova (or star formation) rates 
and higher variance in the yields from different stars will contribute to higher 
cosmic variance in chemical abundances. 

If two metals are produced in the same proportions in supernovae, 
their abundance ratio in the ISM will remain constant. If they are produced at different sites 
or with different proportions, they can be used to  constrain the chemical enrichment 
history of a galaxy, even without any explicit reference to time or, equivalently, 
to the ages of the stars being analyzed.  This is the case, for example, for oxygen and iron, 
with the former being mainly produced in massive stars that die as Type II supernovae, 
and the latter coming chiefly from Type Ia supernovae, the result of the 
evolution of lower mass stars. If determined observationally, the scatter in the abundance
ratios of O/Fe at any given Fe can provide valuable information on 
the star formation rate and nucleosynthetic yields, complementing present-day 
abundance distributions in stars.

For years, efforts to determine the spread in oxygen abundances, 
or [O/Fe]\footnote{We use the
standard bracket notation, [a/b] 
$= \log \frac{\rm N(a)}{\rm N(b)}  - 
\log \left(\frac{\rm N(a)}{\rm N(b)}\right)_{\odot}$,
 where N(x) represents the number density of nuclei of the element x.}, 
at any given [Fe/H] have been unsuccessful. \citet{2006MNRAS.367.1329R} found a spread of 
about 0.07 dex for both thin- and thick-disk members based on abundances from 
the O I triplet at 777~nm that were calibrated to other lines to minimize departures 
from local thermodynamical equilibrium (LTE). \citet{2013ApJ...764...78R}, 
using again the O I triplet and including detailed non-LTE calculations for their sample, 
found a spread of about 0.05 dex among kinematically high-confidence members of the thin or thick disks. 
Most likely, in these and other studies, the observational and analysis uncertainties were at 
the same level as the observed spread in each population.

Recent progress in differential studies of solar analogs has shown that a precision
better than 0.01 dex is possible \citep{2009ApJ...704L..66M, 2009A&A...508L..17R, 
2015arXiv151106583B}. 
\citet{2015A&A...579A..52N} examined the abundances of 14 elements in about twenty nearby
stars with atmospheric parameters very close to solar (within 100~K in $T_{\rm eff}$, 
about 0.1~dex in $\log g$ or [Fe/H]) and found a scatter in [Ca/Fe] or [Cr/Fe] of about 0.01~dex 
 for the thin-disk members in his sample. In contrast, the scatter for other
elements is several times larger and in all cases clearly higher than the measurement
uncertainties. For the $\alpha$-elements Mg, Si, S, and Ti, Nissen 
found a tight correlation between their abundance ratio to iron and stellar age,
inferred from the comparison with models of stellar structure and evolution, and thanks to the extreme accuracy of
the atmospheric parameters provided by the differential analysis relative to the Sun.

These results have revealed the cosmic scatter in the abundance ratios of thin-disk
stars and provide new strong constraints for chemical evolution models of
the solar vicinity. Obviously, it is desirable to extend these measurements beyond the
solar neighborhood, and in particular to carry out similar analyses for thick-disk
stars. As we describe in this paper, the Apache Point Galactic Evolution Experiment
(APOGEE) provides such data.

We present oxygen abundances from OH lines in the H band (1.5--1.7~$\mu$m) 
for disk stars over a wide range of distances observed by APOGEE, which is part of the Sloan 
Digital Sky Survey (SDSS). We overcome systematic effects by considering 
subsamples with similar atmospheric parameters. Within these subsamples, 
we show that oxygen abundances are extremely precise, as quantified by the spread 
observed in open clusters, about 0.005~dex at 4000~K and solar metallicity.
Based on these measurements, for a sample that is roughly 100 times 
larger than in any previous study and extends to the inner parts of the Galaxy, we 
discuss the cosmic scatter 
in the oxygen-to-iron abundances for the thin- and thick-disk stellar populations.

\section{Analysis}

Using the SDSS 2.5m telescope \citep{2006AJ....131.2332G}, 
APOGEE started in 2011 to map the chemical abundances of Milky Way stars, 
with an emphasis on the dust-obscured populations in the central parts of the 
Galaxy and the disk \citep{2015arXiv150905420M}. The latest data release 
(SDSS-III DR12; \citet{2015ApJS..219...12A}, \citet{2015AJ....150..148H}) includes spectra, 
atmospheric parameters, and chemical abundances for some 150,000 stars.
We refer to \citet{2013AJ....146...81Z}, \citet{2015AJ....150..173N}, \citet{2015arXiv151007635G}, 
\citet{2015ApJS..221...24S}, and \citet{2015AJ....149..181Z} for the details of the APOGEE targeting, 
data processing, and analysis pipeline. The observations continue
and will be complemented with a second, basically identical, instrument from 
the southern hemisphere, starting later in 2016 \citep{2012SPIE.8446E..0HW}.

\subsection{Sample selection}
\label{selection}

We based our analysis on the stars in DR12, adopting 
 atmospheric parameters and abundances included in that data release\footnote{DR12 Summary table `allStar-v603.fits'}. 
\citet{2014ApJ...796...38N} examined the spread in [$\alpha$/Fe] for stars in the
thin disk as a function of the median signal-to-noise ratio per pixel 
($S/N$) in the DR10 APOGEE spectra\footnote{We refer to the visit-combined 'apStar' APOGEE spectra, with
roughly three pixels per resolution element.} \citep{2014ApJS..211...17A}, finding that the value flattened at 
about 0.025 dex for $S/N>300$ (most APOGEE targets have been observed to $100<S/N<200$). 
The APOGEE Atmospheric Parameters and Chemical Abundance Pipeline
(ASPCAP; \citet{2015arXiv151007635G}) determines an 'average' $\alpha$-element
to iron ratio simultaneously with the effective temperature, surface gravity, metallicity,
and the abundances of carbon and nitrogen. This $\alpha$-abundance is derived
by varying in block the abundances of O, Mg, Si, S, Ca, and Ti. At low
effective temperature $T_{\rm eff}< 4500$ K, the inferred [$\alpha$/Fe] follows
mainly [O/Fe] due to the many OH lines in the H band, but with
contributions from atomic transitions of the other $\alpha$-elements. 

Instead of [$\alpha$/Fe], we examined the [O/Fe] values derived from OH lines 
exclusively (the `calibrated' values provided as part of the ELEM array of
abundances in DR12 -- see \citet{2015AJ....150..148H} for details). Our [Fe/H] values
are from the corresponding calibrated values in the PARAM array. In an attempt 
to retain the maximum precision avoiding systematic errors, we separated stars into 
different $T_{\rm eff}$ intervals and analyzed each bin independently.

We selected stars in the APOGEE sample observed with the SDSS 2.5m telescope
 that comply with the following requirements: $4000<T_{\rm eff}<4600$~K, 
 $-0.65<{\rm [Fe/H]}<+0.25$~dex, $S/N>80$, $\chi^2_{red}<25$, 
and since we rely on the \textup{{\it \textup{calibrated}} }ASPCAP parameters, 
only low-gravity stars were selected ($\log g < 3.8$). 
The lowest effective temperature was set at 4000 K to avoid problems 
with the oxygen abundances from APOGEE for cooler stars \citep{2013AJ....146..133M, 2015AJ....150..148H}. 
In addition, we avoided bulge stars by restricting the range of projected
galactocentric radii to $R_{g}>3$~kpc and with $\vert{z}_{g}\vert<2.5$~kpc (distances from \citet{2015ApJ...808..132H}). 
After applying these filters, our sample contains 16,870 stars.

\subsection{Chemical split between the thin and thick disk}
\label{split}

\begin{figure*}[ht!]
        \centering
        %\resizebox{\columnwidth}{!}
        {\includegraphics[width=15cm]{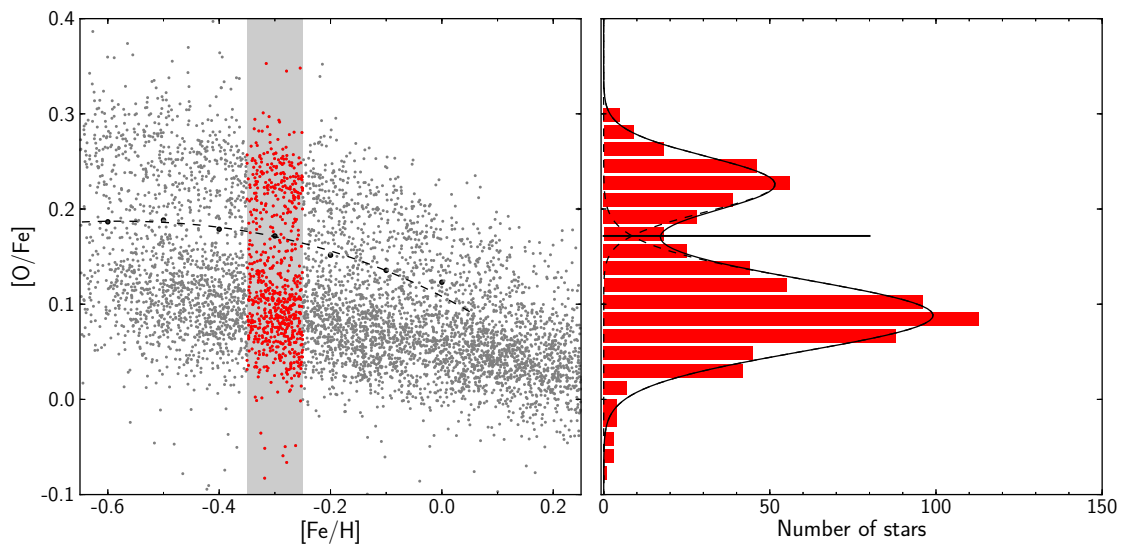}}
        \caption{Example of the procedure used to chemically separate thin and thick 
        disks by fitting a double Gaussian to the [O/Fe] distribution. Data correspond 
        to the bin 4200$<T_{\rm eff}<$4400~K. In the right 
        panel the [O/Fe] distribution for the bin -0.35$<$[M/H]$<$-0.25~dex is shown. 
        The separation between disks is marked with dots on the left panel 
        and with a straight line between the Gaussians for the corresponding [M/H] bin on the right side.}
        \label{ChemSep}
\end{figure*}

Stars in the thin and thick disks can be statistically distinguished based on their
distance to the plane, age, kinematics, or chemical compositions. 
The kinematic properties of both disks overlap, and the 
  available distances are not very accurate. On the other hand, the intrinsic 
uncertainties in the determination of abundances by the ASPCAP pipeline are very small, 
and therefore we prefer to use chemistry to distinguish the membership
of stars in one of the two disk populations. From all the 
$\alpha$-element abundances provided by ASPCAP, 
we selected oxygen because of the 
large number of OH lines detected in the APOGEE spectra.

We split our sample into three bins in  temperature and nine bins in metallicity, with widths of 
200~K and 0.1~dex, respectively. The number of stars in each bin ranges
from 210 to 1100. 

Stars in the thin and thick disk split quite nicely using [$\alpha$/Fe] from the APOGEE
measurements \citep{2014A&A...564A.115A, 2014AJ....147..116H, 2014ApJ...796...38N}. 
To separate the two components, we used a least-squares fit with
a double Gaussian to the [O/Fe] distribution from each bin, as illustrated in Fig. 1. 
The standard deviation of each Gaussian is taken as the measured
scatter (cosmic plus 
measurement uncertainties). This method provides
reliable estimates of the scatter, even when the separation between disks is not clear 
and they overlap, as is usually the case. In addition, we can define a chemical division between 
the disks in the intersection between the two Gaussians, except at high [Fe/H] when the two
populations completely overlap (see Fig.~\ref{ChemSep}). 

The uncertainty for each bin in the [O/Fe] histogram (see the right-hand panel 
in Fig. \ref{ChemSep}) was taken as the square root 
of the number of stars in the bin. These were considered when fitting double Gaussians. 
The square root of the diagonal of the inverse of the curvature 
matrix provides the errors in the mean and $\sigma$ for both Gaussians.

\subsection{Estimating the [O/Fe] uncertainties}
\label{sigmaerr}

To measure the cosmic dispersion of a certain element, we need to not only 
measure the spread, but also properly estimate the actual uncertainties in 
our abundance measurements. The expected errors in abundance measurements are often
underestimated because it is difficult to account for all relevant contributions.  
For this reason, we did not attempt to  calculate the uncertainties 
in the oxygen abundances, but instead derived them empirically. Thanks to the wide sky coverage 
of the APOGEE survey, multiple clusters are included in our sample 
\citep{2013ApJ...777L...1F, 2015AJ....149..153M}. 

We assumed that the cosmic variance in a cluster 
is negligible, and therefore the measured scatter mainly reflects our 
analysis uncertainties. \citet{2016ApJ...817...49B} has recently argued that the APOGEE data 
can be used to set tight limits to the intrinsic spread in the abundances 
of carbon and iron in open clusters at $\lesssim 0.01$, and at $\lesssim 0.015$ for oxygen,
which supports this argument. Globular clusters show multiple populations and abundance 
anomalies, and therefore the measured spread in these systems provides only upper limits
to the intrinsic uncertainties, but even these are useful for our purposes.

We selected clusters from the list of APOGEE calibration clusters with 
metallicity above $-1.3$ dex: M5, M67, M71, M107, NGC 2158, NGC 6791, and NGC 6819. 
Stars observed in the field of each cluster were first selected as potential members 
according to their coordinates (stars within the cluster tidal radius), radial velocity (RV) 
in the range of $\pm 30$ km/s around the mean cluster value, and avoiding dwarf stars. 
The mean cluster RVs were taken from the SIMBAD astronomical database \citep{2000A&AS..143....9W}, 
and the cluster radii are the same as those used in M{\'e}sz{\'a}ros et al. 2013. 
The final sample retains stars  within a 2$\sigma$ interval
around the mean [Fe/H] and RV of the potential members.

\begin{figure}[t!]
        \centering
        \resizebox{\columnwidth}{!}
        {\includegraphics[trim=0 15 0 05, clip]{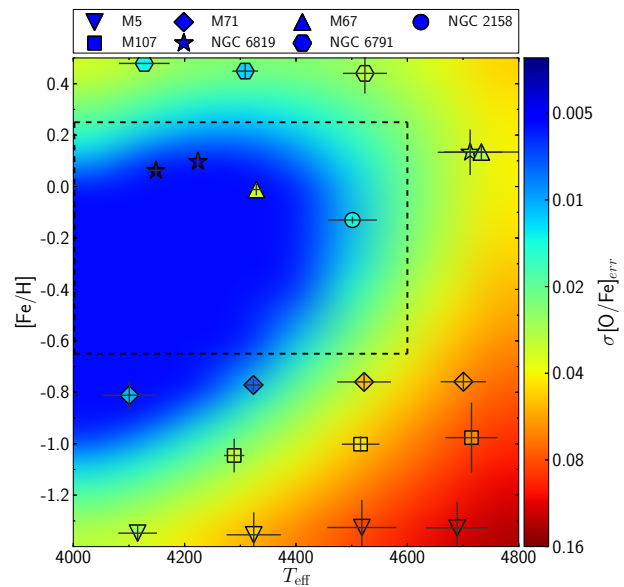}}
\caption{Spread in the ratio of oxygen to iron abundances measured in different clusters 
  as a function of the mean $T_{\rm eff}$ and [Fe/H]. The background color indicates the 
  variations predicted by the polynomial fit to the data (capped to have minimum value of 0.005 dex), 
  and follows the same color code as used for the individual data points. The  error bars shown for 
  the data points reflect the standard deviation. The stars from our sample 
  lie in the region marked with dashed lines.}
\label{Poly}
\end{figure}

We explored a broader range in [Fe/H] and $T_{\rm eff}$ than in our sample (Sect. \ref{selection})
to better map the sensitivity of the uncertainties to these parameters.
For each cluster we divided the stars into $T_{\rm eff}$ bins and measured the
[O/Fe] rms scatter in those with more than one star. Since the stars 
in some clusters span multiple bins, we have more data points than
clusters. The results
are shown in Fig. \ref{Poly}. We find that the scatter is a smooth function of 
effective temperature and metallicity and that the polynomial
\begin{eqnarray}
\sigma[{\rm O/Fe}]_{\rm err} = &  1.915       & - ~ 9.143\times 10^{-4} T_{\rm eff}  \nonumber \\
                  + &  2.837\times 10^{-1} {\rm [Fe/H]}& - ~  6.108\times10^{-5}  T_{\rm eff} {\rm [Fe/H]} \nonumber \\
                  + & 1.091\times 10^{-7} T_{\rm eff}^2  & + ~ 4.119\times 10^{-2} {\rm [Fe/H]}^2  \nonumber
\end{eqnarray}

\begin{figure*}[]
        \centering
        {\includegraphics[width=15cm]{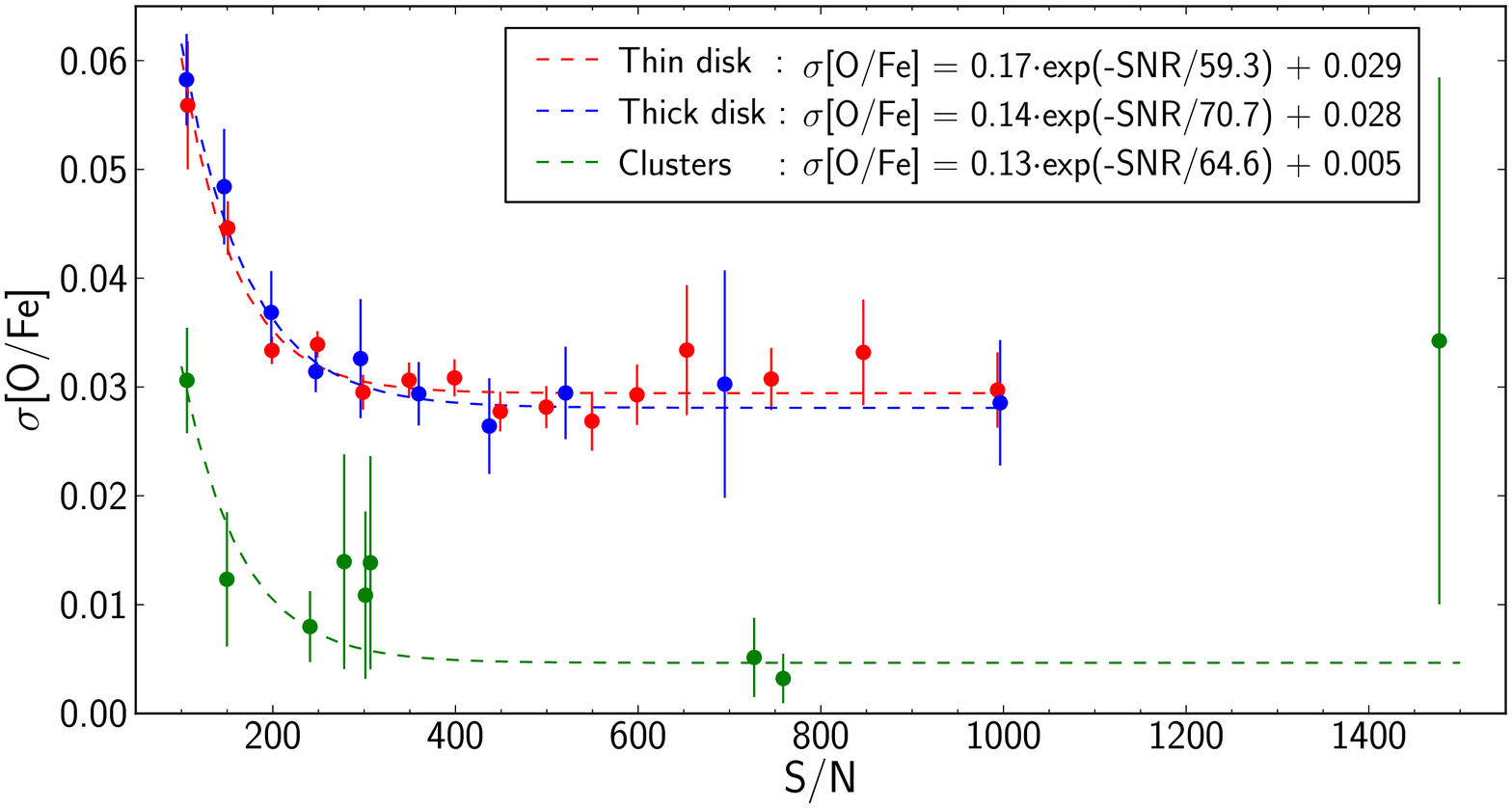}}
\caption{1-$\sigma$ standard deviation in the ratio of oxygen to iron abundances 
  as a function of the S/N for stars with 4200$<T_{\rm eff}<$4600~K 
  and 0.05$<$[Fe/H]$<$-0.45~dex. An exponential is fitted to the data points, 
  weighted with the corresponding uncertainties. The point at very high $S/N$ 
  corresponds to the M67 bin that contains a star whose [O/Fe] is remarkably 
  different from the cluster average (see Sect. \ref{sigmaerr})}
\label{ScatterSNR}
\end{figure*}

\noindent closely follows the data with an rms scatter of 0.01 dex, 
but we capped the polynomial predictions ($\sigma$[O/Fe]$_{\rm err}$) to avoid values 
lower than 0.005 dex, the average of the three lowest data points in our measurements.
The M67 data point at $T_{\rm eff}\sim$~4300~K is the only one that clearly 
deviates from the fitted polynomial. This is due to a star, 2MASS J08493465+1151256, 
whose RV and [Fe/H] values agree with the cluster mean, but which shows a remarkable 
difference of 0.09~dex in [O/Fe] with respect to the mean value of the cluster. 
Nevertheless, we decided to keep this star and avoid exceptions to our cluster 
selection criteria, since it has a negligible effect on our derived polynomial.
In addition, we did not discard the possibility that M67 might be chemically inhomogeneous 
at the level of $\sim$0.02~dex, as has previously been found in other open clusters 
\citep{2016MNRAS.457.3934L}.

To ensure that the cluster sample is representative 
of the whole sample, we also examined the $S/N$\footnote{We define $S/N$ as the
median value for a given spectrum. The reported values tend to be optimistic
for $S/N \gtrsim 300$, which in this range is limited by 
imperfections in the behavior of IR detectors.} of the data. 
While the mean $S/N$ of our sample is 295,  clusters have on average 
lower $S/N$ values,  except for M67 ($S/N\sim$~545), 
NGC~6819 ($S/N\sim$~324), and NGC~2420 ($S/N\sim$~316). 
The remaining clusters have a $S/N$ between 
150 and 200. We study this matter in more detail below.

\subsection{[O/Fe] dispersion against signal-to-noise ratio}
\label{snr}

\citet{2014ApJ...796...38N} have shown  that the empirical scatter in [$\alpha$/Fe] 
for the low-$\alpha$ stars in APOGEE decreases exponentially as a function of $S/N$, 
reaching a plateau of $\sim$0.023~dex at $S/N\gtrsim$200.

After we separated between thin- and thick-disk stars in our sample, we calculated the 
dependence of the standard deviation in [O/Fe] on $S/N$. We selected stars with 
4200$<T_{\rm eff}<$4600~K and 0.05$<$[Fe/H]$<$-0.45. Lower temperatures 
or higher metallicities were avoided to guarantee a minimum number 
of 25 stars per bin for both the thin and thick disks. Following the same method used to calculate
the scatter and split the disks, we divided our sample into bins of 
$T_{\rm eff}$ and $S/N$. For each bin we fit a third-order polynomial 
to the [O/Fe] vs. [Fe/H] trend to subtract it, and then fit a Gaussian 
profile to the [O/Fe] histogram. The scatter in [O/Fe] was taken as the 
standard deviation of this Gaussian. Finally, we averaged the scatter 
found for different temperatures at the same $S/N$.
As can be seen in Fig. 3,  the dispersion
in [O/Fe] does indeed depend on $S/N$. Although we do not reach  a  dispersion as small 
as \citet{2014ApJ...796...38N} for the thin disk, the trends are very similar.

We  also show data for the clusters, avoiding 
those with parameters that are very different from those in our sample. 
The scatter in this case is calculated as the standard deviation of 
[O/Fe] for each bin since we do not have enough stars per bin to perform a Gaussian fit. Our cluster sample 
spans a wide range in $S/N$, but their dispersions are still remarkably 
lower than those for field stars. We therefore conclude that our estimates of
the measurement uncertainties from clusters are not  
underestimated because of the differences between clusters and sample $S/N$. 
We note, nonetheless, that the  uncertainties in the  cluster data are larger 
than in the sample data as a result of the limited number of stars per cluster 
for each bin.

\section{Results and conclusions}  

\begin{figure}[]
        \centering
        \resizebox{\columnwidth}{!}{\includegraphics[trim=0 40 0 65, clip]{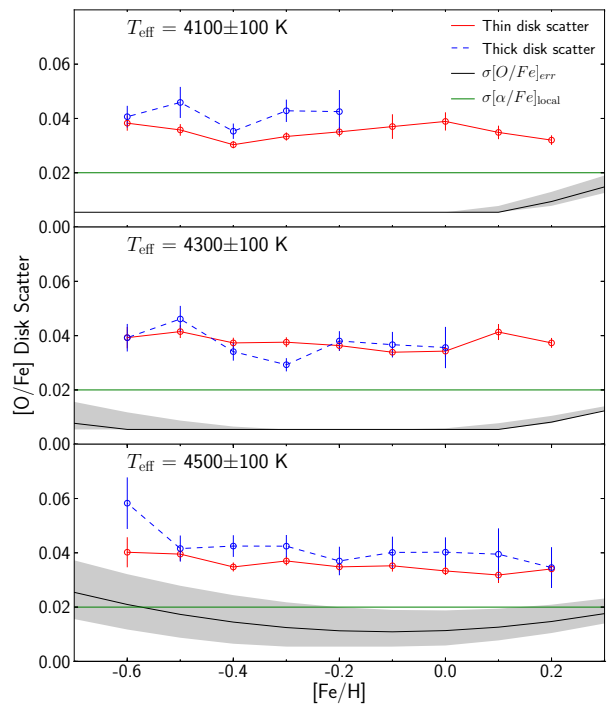}}
\caption{1-$\sigma$ scatter measured in the thin (red solid curve) and thick 
(blue dashed line) disks. 
The black line shows the estimated uncertainties in the APOGEE [O/Fe] ratios 
from the dispersion 
empirically measured in open clusters, while the gray area indicates the size 
of the variation in 
these uncertainties between the extreme temperature values in each temperature bin.}
\label{Scatter}
\end{figure}

Figure \ref{Scatter} shows the 1$-\sigma$ scatter measured for the
thin- (red) and thick-disk (blue) populations in each temperature
and metallicity bin. The black line shows the estimated 
uncertainties in the APOGEE [O/Fe] ratios from the dispersion 
empirically measured in clusters, while the gray area indicates
the variation between 
extreme temperature values in each temperature bin.
The estimated uncertainties reach a minimum value of 0.005 dex 
for solar-metallicity stars with $T_{\rm eff}\simeq 4000$ K and increase 
for warmer or more metal-poor stars.

The APOGEE measurements unambiguously detect  the cosmic scatter
in [O/Fe] at any given metallicity.  
The [O/Fe] scatter is not very different in the thin and thick disks,
and the disks show an approximately flat value between 0.03--0.04 dex 
over the entire range of metallicity considered.

Our APOGEE sample includes disk stars at distances of up to 30~kpc from
the Sun (75\% of the sample are within $6<R_{g}<12$~kpc), 
and therefore it is interesting to compare our derived
scatter in [O/Fe] with the value for the thin disk from \citet{2015A&A...579A..52N}, 
which was based on solar twins within $\sim$ 40~pc. Nissen's oxygen
abundances are based on a single feature, the weak forbidden line at 630 nm,
which is blended with a Ni I transition, and this is likely to cause part of 
the scatter in [O/Fe] he reported. We therefore chose to compare our results with the
scatter he reported for other $\alpha$ elements that show a similar
correlation between abundance and age, namely Mg, Si, S, or Ti. In these,
the scatter in [X/Fe] is 0.022, 0.016, 0.021, and 0.017 dex, respectively. 
Taking $\sigma {\rm [\alpha/Fe]}_{\rm local}=0.02$ dex as a representative value, 
marked in Fig. \ref{Scatter} with a green solid line, 
the APOGEE [O/Fe] ratios show nearly twice as much scatter as the value
found in the immediate vicinity of the Sun.

\citet{2015A&A...579A..52N} found that a significant fraction of the scatter 
in [Mg/Fe], [Si/Fe], [S/Fe], or [Ti/Fe] among 
thin-disk stars in the solar neighborhood can be attributed to chemical evolution.
These abundance ratios correlate tightly with  stellar age, and when the
age trend is removed, the scatter reduces to $\sim 0.01$ dex, or about half
of that before removing the trend. Obviously, this contribution must
be present in the APOGEE sample, but since the volume probed by APOGEE 
observations is much larger, additional effects are likely to 
contribute to the scatter. 

The abundance of [Fe/H] or the alpha elements [$\alpha$/H] decreases significantly 
with galactocentric distance for the thin disk at about 0.05 dex/kpc, 
while variations are much smaller for the thick disk \citep{2014AJ....147..116H}. 
On the other hand, at any given [Fe/H], the changes  in [O/Fe] as a function 
of galactocentric distance are very small for both populations. If we use
the APOGEE data to evaluate how much the mean radial and vertical gradients 
contribute to the scatter, we find that its contribution is negligible.

As a result of radial migration,  stars currently in the solar neighborhood were formed
at a wide range of galactocentric distances, between 2 and 13 kpc, with about 75\%
of them formed at galactocentric distances between 3 and 9 kpc \citep[e.g.,][]{2013A&A...558A...9M}. 
This range, however, is not much larger than that
spanned by the stars in the APOGEE sample, but the latter have also been subjected 
to radial mixing and therefore sample a larger volume of the disk. Spatial variations 
in the star formation rate or in the gas outflow rate could be responsible 
for an increased scatter.
Indeed, limiting our APOGEE sample to stars at less than 1~kpc from the Sun,
we measure a scatter of $\sim$0.03~dex in the solar neighborhood. 
On the other hand, stars at more than 4~kpc from the Galactic plane 
show a scatter that  rises to $\sim$0.1~dex,
suggesting that the dispersion in [O/Fe] is larger in the Milky Way halo 
than in the disk.

The potential use of information on the spread in abundance ratios at any given metallicity
can be illustrated with a back-of-the-envelope calculation. With an average
expansion speed of $10^3$ km s$^{-1}$ and a time for the ejecta to merge into
the ISM of about $10^4$ years, a supernova remnant can pollute a region
$\sim 10$ pc in radius\footnote{As an example, the Crab nebula, the remnant of a supernova explosion in 1054, 
is currently expanding at about 900 km s$^{-1}$ 
and has a radius of about 3 pc.}. Assuming a typical mass of metals in  the ejecta of 
$\sim$ 1 M$_{\odot}$, 
a single supernova increases the density of metals in the surrounding ISM
by $\sim 5 \times 10^{26}$ kg pc$^{-3}$. A maximum density for the ISM where stars form
is given by the typical density for a molecular cloud, about $5 \times 10^{32}$ kg pc$^{-3}$,
and therefore the first Galactic supernova would bring the metallicity 
of the surrounding ISM  to [Fe/H]$\sim -3$. 

The variance in metallicity across individual zones, $\sim$10~pc in size,
will be large in the early Galaxy, but as more and more supernova 
explode, this variance will be progressively reduced. Considering the metallicity
distribution at the present time involves measuring the variance both spatially (across zones) and
over time. Considering the spread in abundance ratios at any given metallicity 
would be more similar to evaluating the variance across zones at a particular 
moment in time, but we note that this is a rough approximation
because the age metallicity relationship in the Milky Way disk is only loosely
defined \citep[e.g.,][]{1993A&A...275..101E, 2013A&A...560A.109H}.

Based on these order-of-magnitude arguments and assuming that low-mass stars form 
at a constant rate, we can show numerically that a constant
supernova rate of about 0.01 to 0.1 per 10-pc zone per million years would lead, 
after 10$^{10}$ years, to a metallicity distribution with a 1$-\sigma$ dispersion 
across zones and over all ages of about 0.2 dex. This supernovae rate would at the same time produce
a dispersion about ten times smaller over zones at any given moment in time in the last
$\sim 5\times10^9$ years. These figures are consistent with the spread in the metallicity distribution 
in the thin or thick disks (about 0.2 dex) and with the spread in oxygen abundances
that we find at any given metallicity (0.02--0.04 dex).

We can further support this claim by assuming that the Galactic disk is a flat cylinder
30~kpc in diameter and 1~kpc in height. The cylinder would contain some $10^8$
zones of 10 pc each\footnote{The volume of the Galactic cylinder would be 
$7\times10^{11}$ pc$^3$, which divided by the volume of the sphere polluted by a supernova 
remnant, $4\times10^3$ pc$^3$, gives us the number of zones.}. 
A supernova rate of 0.01 per zone per million years implies $n= 10^{10}$ 
supernovae in the entire Galactic disk in $10^{10}$ yr. The events would be distributed
following a binomial distribution with a mean of $10^{10} \times 10^{-8}= 10^2$ 
supernovae per zone and a standard deviation of $\sqrt{10^{10}\times 10^{-8} 
\times (1-10^{-8})} \simeq 10$, which leads to a spread in metals of 10/100/ln(10.) = 0.04 dex.

These simple arguments show that the star formation rate must have been
much higher in the past than it is now -- one supernova per century, the current
rate, is roughly equivalent to $10^{-4}$ supernova per 10-pc zone per million years.
A quantitative discussion requires more sophisticated models, but we conclude that  
as the precision of our abundance analyses reaches and  surpasses 0.01 dex, 
new diagnostics will become available to constrain models of the formation and 
evolution of the Milky Way. 

\begin{acknowledgements}
We thank the anonymous referee for his/her thorough review and suggestions, 
which significantly contributed to improving the quality of the publication. 
We are also grateful to I. Minchev and J. Bovy for useful comments.
S.B. and C.A.P. acknowledge financial support from the Spanish Government through the 
project AYA2014-56359-P. P.M.F. is supported by NSF grant AST-1311835.
S.M. has been supported by the J{\'a}nos Bolyai Research Scholarship of the Hungarian Academy of Sciences.
Funding for SDSS-III has been provided by the Alfred P. Sloan Foundation, 
the Participating Institutions, the National Science Foundation, and the U.S. 
Department of Energy Office of Science. The SDSS-III web site is http://www.sdss3.org/.
SDSS-III is managed by the Astrophysical Research Consortium for the Participating 
Institutions of the SDSS-III Collaboration including the University of Arizona, 
the Brazilian Participation Group, Brookhaven National Laboratory, 
Carnegie Mellon University, University of Florida, the French Participation Group, 
the German Participation Group, Harvard University, 
the Instituto de Astrofisica de Canarias, the Michigan State/Notre Dame/JINA 
Participation Group, Johns Hopkins University, Lawrence Berkeley National Laboratory, 
Max Planck Institute for Astrophysics, Max Planck Institute for Extraterrestrial Physics, 
New Mexico State University, New York University, Ohio State University, 
Pennsylvania State University, University of Portsmouth, Princeton University, 
the Spanish Participation Group, University of Tokyo, University of Utah, 
Vanderbilt University, University of Virginia, University of Washington, 
and Yale University. 

\end{acknowledgements}

\bibliographystyle{aa} % style aa.bst
\bibliography{OFeCosVar} % your references Yourfile.bib

\end{document}